\documentclass{article}
\usepackage{e}
\begin{document}
\branch{A}
\DOI{123}                       
\idline{A}{1, 1--11}{1}         
\editorial{}{}{}{}              
\title{LHC Symposium 2003: Summary Talk}
\author{J. A. Appel\
}
\institute{Fermilab, PO Box 500, Batavia, IL  60510, USA\\
e-mail: appel@fnal.gov}
\maketitle
\begin{abstract}
This summary talk reviews the LHC 2003 Symposium, focusing on expectations 
as we prepare to leap over the current energy frontier into new territory.
We may learn from what happened in the two most recent examples of leaping 
into new energy territory.  Quite different scenarios appeared in those 
two cases.  In addition, we review the status of the machine and 
experiments as reported at the Symposium.  Finally, I suggest an attitude 
which may be most appropriate as we look forward to the opportunities 
anticipated for the first data from the LHC.
\end{abstract}

As we contemplate the three days of excellent talks we have just 
experienced, we are invited to think about how to convey our science and 
its goals to the public.  In that context, we should understand where the 
public perceptions are.  I am reminded of a recent discussion among 
knowledgeable people, motivated by the book ''The End of Science" by John 
Horgan.  In it, Horgan says ''And now that science - true, pure, empirical 
science - has ended, what else is there to believe in?"  It is too bad 
that anyone thinking this was not here at this Symposium!  We are here 
reaffirming that empirical science is alive and well.  

\section{Leaping Over Energy Frontiers}

Even more than simply continuing the empirical research of the past, we 
are at the threshold of a new era, with a new leap beyond the current 
energy frontier.  Following the excellent presentations at this Symposium,  
it is perhaps worthwhile to pause a moment and consider our most recent 
leaps of energy frontiers.  What do they suggest?

What happened when the ISR and the "200 GeV" machine turned on?  Available 
center-of-mass energy jumped from ~8 GeV to  20-50 GeV.  New energy 
territory opened to us.  We were surprised, even shocked by how different 
the world seemed.  Almost immediately, we saw the advent of high-$p_t$ 
events at both the CERN ISR and at Fermilab.  Pions were observed with 
cross sections no longer dropping exponentially with 
$p_t$.\cite{cronin,busser}  
Rather, the drop with $p_t$ was more like a power-law, eventually reaching 
that for hard point-like scattering!  Backgrounds for many planned 
experiments were orders of magnitude larger than expected.  More 
fundamentally, we observed (as we now understand it) the effects of the 
quark substructure of hadrons.  

We also started to produce particles essentially undreamed of before - 
well, dreamed of by only a few foolhardy visionaries.  In addition to the 
pions at high $p_t$ coming from hadronic interactions, a plethora of 
leptons appeared.  Their numbers could not be explained by the decay of 
known strongly-produced particles.\cite{bourquin}  Eventually, these 
leptons were seen to come from the semileptonic decays of the 
previously-unknown heavy quarks.

Perhaps the excess of leptons reminds you of the apparent excesses of 
heavy quarks seen in hadronic interactions today (especially of B mesons 
and $J/\psi$ and $\psi'$ onia).  It may be that what Mary Bishai referred 
to as a $b$-production excess of 1.2-1.9 times theory,\cite{bishai} will 
continue to fall as theoretical models of production are refined.   The 
fractional excess does seem to be coming down with time.  However, it is 
possible that we are already seeing the effects of something which we will 
only understand once we have data from the LHC.

What happened when the big CERN and Fermilab hadron colliders turned on?
Available energy jumped from a few tens of GeV to 630 and 2,000 GeV.  
Again, new 
energy territory opened for exploraton.  We were again surprised - maybe 
not so much by a new energy scale which was predicted ($W$ and $Z$ 
masses), but by the very large mass of the top quark.  I remember well, 
how upon seeing evidence for the bottom quark, we immediately expected to 
see the top quark at ~$\pi$ times the mass of the bottom quark, just like 
the factor between the bottom quark and the strange quark.  The ratio of 
top to bottom quark masses is more like 40 than 3!   We do not understand 
why the top quark is so heavy to this very day.  

We have seen no direct evidence of any of the suggested new particles: 
not sequential $W$ or $Z$ bosons, not Higgs, not SUSY, nor 
techni-particles.  We have not seen a break in $p_t$ spectra, nor the 
onset of a new level in the hierarchy of matter, nor any suggestion of 
something more fundamental than quarks and leptons.

\section{How the Preparations are Going}

As you have shown at this symposium, you are building detectors, and 
solving 
technical and managerial problems.  You are also building expanded 
collaborations and new tools to deal with the new sociology: learning
how to live with larger and increasingly internationalized collaborations, 
learning new techniques and tools for ever larger projects, 
and beginning to experiment with new computing paradigms like GRID.
\cite{lothar,brook}

I have been impressed by the trigger tables shown and the expanding 
physics goals shown by experiments.  We had talks on heavy-ion collision 
measurements in the big $p-p$ detectors, ATLAS and CMS, detection of 
quark jets in ALICE, and the appearance of B physics everywhere.  
Detectors have had design and engineering updates, and simulations 
continue to include more complete detector modeling.  As usual, the 
results suggest somewhat less capability, but hopefully more realistic 
expectations.  At the same time, perhaps motivated in part by the new 
understanding, better algorithms have been developed to compensate for 
somewhat reduced detector expectations; e.g., in tracking and and heavy 
quark tagging algorithms.  In order to continue this progress, mock data 
challenge efforts cannot be over-valued, both for improving the physics 
reach of experiments and and for debugging the computing environment of 
the future.  Even more, better motivation will come from the data itself 
once you have the real thing.

As an example of how time with actual physics data helps, let me cite the 
work reported by Juan Estrada at a seminar at Fermilab just the week 
before the Symposium, and referred to by Jianming Qian.\cite{qian}  Unlike 
previous CDF and DZero top-quark mass analyses that used templates, this 
new DZero analysis uses lepton plus jets events and makes a direct 
calculation of the signal and background probability for each event.  That 
probability depends on all measured momenta of the final state lepton and 
jets, and each event's contribution depends on how well it is measured.  
The quoted preliminary result for the top quark mass is $M_t = 180.1 \pm 
3.6 \pm 4.0 GeV$.  The improvement in statistical error is equivalent to 
a factor of 2.4 in the size of the data sample.  The relative error in 
this one decay channel alone is 3\%, compared to 2.9\% from the previous 
combined CDF and DZero average for all analysed decay channels.

\section{Progress, Yet Concerns}

You have shown real, substantial progress from the past year at this 
Symposium.  It has been very good to see the progress on the LHC itself, 
and on the detectors, software, and physics planning.  We can all be 
happy that civil construction is now going well, and magnet production is 
getting better.  Roger Cashmore spoke of the ''nightmare" of the civil 
construction problems that are now behind us.  We are also happy to see so 
many detector components getting into construction.  We have heard about 
facing real challenges.  Some have been technical; e.g., in military 
radiation-hard electronics,  some electronic noise and yield issues, 
material budgets, and radiation damage effects.  Some challenges have been 
financial in origin, leading to scope changes and, sometimes, to 
additional funding.  Other challenges have been with schedules, requiring 
continuous review and adjustments (e.g., lack of test-beam availability).
Personally, I am happy to see  some full system tests, and indications 
that planning for commissioning is getting serious attention.  We are all 
happy to see solutions over the last year to these and other problems .

Your progress is important to us at Fermilab.  First, it is important for 
our physics program (CMS) and our super-conducting magnet program.  Mike 
Witherell, in his Director's welcome to you, noted that only Fermilab's 
Tevatron Collider and neutrino programs are larger here at the Lab.  
Second, your progress is important for the planning of much of the rest of 
our program as well.  For the Tevatron Collider, currently the energy 
frontier machine, the importance is obvious.  However, in fact, your 
progress is important to all of HEP.  Consider the implications for B 
Factories!

Nevertheless, even as an LHC outsider, I have concerns.  The scale of the 
industrial technology needed for the machine and for the detectors is 
still new to our community.  It is not obvious that accelerator components 
will stay ahead of the "just in time" schedule.  We have all been invited 
to look at the CERN LHC "dashboard" on the web.\cite{dashboard}  There 
is no real evidence yet of the rapid change in delivery slopes needed to 
meet the schedule for beams.  

Timing can be everything.  Staging of detector components may get us to 
the point where we cannot buy needed components later.  This is already a 
problem.  Some commercial technologies may not last long enough for our 
development and construction schedules.  DMILL radiation hard ASIC 
technology is going away already; will 0.25 micron ASIC technology be far 
behind?  Even networking and computing components are a worry.  Consider 
the "Objectivity" software suite.  At the same time, there are more 
technology decisions yet to be made than is healthy at this stage.  I 
would mention the CMS pixel size, the ATLAS B-layer pixel size, the CMS 
electromagnetic calorimeter electronics, and the LHCb RICH 
photon-detection decision especially.

Common computing approaches can save duplication, and help by stressing
systems in more than one environment.  This can head off problems later, 
as the environments evolve for all those using a given approach.  Yet, 
this commonality of approach is just starting.  I was surprised in the 
talks of Lothar Bauerdick\cite{lothar} and Nick Brook\cite{brook} to see 
so far how little integrated into the big experiments are the GRID 
projects.

Testing and commissioning times are getting squeezed almost everywhere - 
already!

\section{A Few Words About the Physics}

The main message from the experiments at this Symposium has been that 
discoveries may be made very early, if nature is as expected by most.  In 
my summary, I have chosen to show some of  
the most frequently referenced transparencies.  Since these plots have 
been shown so often, I won't even have to tell you what they are, even if 
none of us remembers or cites where the plots were first shown!  The
famous Higgs sensitivity plot is made to show the early discovery expected  
over the whole range of likely mass.  Recently, including the vector-boson 
fusion process helps in the previously-difficult low-mass region.  For 
SUSY, again we are assured of rapid discovery, the plots showing possible 
discovery up to $1.5 TeV$ even with only "1 day" of data -- OK, it's a 
good day.  

In the area of Heavy Ions, I'd like to mention the particularly good 
review of Heavy Ion Physics at RHIC by Gunther Roland.\cite{roland}  He 
showed a ''consistent {\it description} of the final state," but noted 
that ''we're missing a picture of [the] dynamical evolution" that gets us 
there from the initial conditions.  Many speakers showed where the LHC 
sits on the phase diagram of temperature vs baryonic chemical potential.  
This plot does not do justice in my eyes to the role of the LHC.  The LHC 
is shown in a tiny corner of the plot.  Yet, the ''missing picture of 
dynamical evolution" may require:
\begin{itemize}
\item More dynamic range in kinematic variables                   
\item Longer time for escaping partons to feel effects of quark-gluon 
plasma
\item Larger samples of charm, bottom, and onium                
\end{itemize}
All these features should be available at the LHC.  The table from the 
talk of Russell Betts\cite{betts} shows quantitatively the much higher 
energy densities, multiplicities characteristic of more quark-gluon 
plasma, and the longer times available for the plasma to influence the 
outgoing states.  All these should make the anticipated effects much 
easier to see at the LHC and to understand.

\begin{table}
\caption{Pb+Pb Collistions at the SPS, RHIC, and LHC}
\centering
\begin{tabular}{|c|c c c|}
\hline 
                & SPS(17) & RHIC(200)  & LHC(5500) \\ 
\hline
$dN_{ch}/d\eta$ &  500    &   700      & 3000-8000 \\
\hline
$\epsilon [GeV/fm^3]$ & $\sim 2.5$  & $\sim 3.5-7.5$  & $\sim 15-40 $   \\
$(t_0 = 1 fm/c)     $ &     1       &       2         &     10          \\
\hline
$ V_f [fm^3]        $ & $\sim 10^3$ & $\sim 7 x 10^3$ & $\sim 2 x 10^4 $\\
                      &     1       &       7         &      20         \\
\hline
$\tau_{QGP} [fm/c]$   & $ \le 1 $   &    1.5-4        &     4-10        \\
                      &     1       &       3         &      7          \\
\hline
\end{tabular} 
\label{table1}
\end{table}

Finally, I'd like to focus on two personal-favorite physics topics:
compossitness and extra dimensions.   Signatures for both of these topics 
may come to be manifest in the same way that high-$p_t$ events came to us 
at the ISR and at Fermilab.  Moreover, such signals can appear quite 
early, with subsets of working detectors and the simplest of analyses.

\section{The Message}

Our keynote speaker, Scott Willenbrock, gave us a similar message to mine.
In many ways, physics has never been more exciting.
\begin{itemize}
\item We are about to extend the energy frontier by a factor of 7.
\item We have an excellent model of what we have seen already.
\item However, we know that our model is incomplete, and we have 
detailed predictions which soon can be tested definitively.
\end{itemize}

We are not at the ''end of science," but hopefully at the threshold of 
exciting new science.  What will the new science be?  I really don't know.  
However, personally, I expect we will have major surprises.  I expect
surprises comparable to those when the CERN ISR and Fermilab began.
In the face of the new energy frontier; 
\begin{itemize}
\item be prepared to read out all working detectors;     
\item be prepared for analysis of early, imperfect data;  
\item be prepared for discovery;                              
\item be prepared for surprises in signal and in backgrounds; 
\item and be prepared to think new thoughts!                  
\end{itemize}

\noindent Good luck!

\section{A Word of Thanks}

I would like to end with a word of thanks 
\begin{itemize}
\item to the organizers,           
\item to the support staff,                       
\item and to the speakers (specially those who responded to my 
request for advance word on their presentations)          
\end{itemize}
for making this such an informative and invigorating symposium.

\end{document}